\documentclass[12pt]{article}
\usepackage{amsmath,amssymb,epsfig}

\usepackage{color}
\input{colordvi.tex}
\def\unit{{\relax{\rm 1\kern-.26em I}}}

\newcommand{\tr}{{\rm Tr}}

\usepackage{bm}
\newcommand{\Slash}[1]{{\ooalign{\hfil/\hfil\crcr$#1$}}}

\makeatletter

\renewcommand\section{\@startsection {section}{1}{\z@}%
                                   {-3.5ex \@plus -1ex \@minus -.2ex}%
                                   {2.3ex \@plus.2ex}%
                                   {\normalfont\large\bfseries}}

\renewcommand\subsection{\@startsection{subsection}{2}{\z@}%
                                     {-3.25ex\@plus -1ex \@minus -.2ex}%
                                     {1.5ex \@plus .2ex}%
                                     {\normalfont\normalsize\bfseries}}

\newcount\hour \newcount\minute
\hour=\time \divide \hour by 60
\minute=\time
\def\now{%
\ifnum \hour<13
  \ifnum \hour=0 \advance \hour by 12 \number\hour:\else \number\hour:\fi%
     \ifnum \minute<10 0\fi%
     \number\minute%
\ A.M.%
\else \advance \hour by -12 \number\hour:%
  \ifnum \minute<10 0\fi%
  \number\minute%
  \ P.M.%
\fi%
}

\makeatother

\begin{document}

\baselineskip=18pt  
\numberwithin{equation}{section}  
\allowdisplaybreaks  



%
%


\thispagestyle{empty}

\vspace*{-2cm}
\begin{flushright}
\end{flushright}

\begin{flushright}
KYUSHU-HET-143
\end{flushright}

\begin{center}

\vspace{1.5cm}

{\bf \Large Radiation of Supersymmetric Particles }

\vspace*{0.2cm}

{\bf \Large  from Aharonov-Bohm R-string}

\vspace*{0.2cm}

\vspace{1.3cm}

{\bf
Yutaka Ookouchi}$^{1,2}$ {\bf and Takahiro Yonemoto}$^{2}$
\vspace*{0.5cm}

${ }^{1}${\it Faculty of Arts and Science, Kyushu University, Fukuoka 819-0395, Japan  }\\

${ }^{2}${\it Department of Physics, Kyushu University, Fukuoka 810-8581, Japan  }\\

\vspace*{0.5cm}

\end{center}

\vspace{1cm} \centerline{\bf Abstract} \vspace*{0.5cm}
We study radiation of supersymmetric particles from an Aharonov-Bohm string associated with a discrete R-symmetry. Radiation of the lightest supersymmetric particle, when combined with the observed dark matter density, imposes constraints on the string tension or the freeze-out temperature of the particle.  We also calculate the amplitude for Aharonov-Bohm radiation of massive spin $3/2$ particles.

\newpage
\setcounter{page}{1} 



\section{Introduction}

A discrete symmetry is a useful tool to construct a realistic model in particle physics (for example, see \cite{refRsymmetry,examples2} and references therein). In particular, R-symmetry is one of the important ingredients in supersymmetric model building, since the lightest supersymmetric particle stabilized by the symmetry offers a natural candidate for the dark matter. 

Recently, Banks and Seiberg gave a new argument on discrete symmetries from a viewpoint of a quantum theory of gravity \cite{BanksSeiberg}. They extended the so-called ``no global symmetries theorem'' to include discrete symmetries. Accepting their arguments, one would be lead to an interesting avenue for string phenomenology: A discrete symmetry, if not broken in coupling to gravity, have to be gauged. Banks and Seiberg also showed the universal effective Lagrangian of a discrete gauge theory by means of the BF coupling. In the effective theory, in addition to a massive gauge field, there is a Kalb-Ramond 2-form field which naturally couples to a string-like object, so-called Aharonov-Bohm (AB) string. Also, there is a particle, called Aharonov-Bohm particle, coupling to the massive gauge field. This is the other ingredient of the discrete gauge theory. As in the well-know Aharonov-Bohm effect for a solenoid \cite{Original}, AB strings and AB particles have quantum mechanical interactions. As was firstly pointed out in \cite{AlWil}, a moving AB string radiates AB particles by the interaction. Explicit calculation of the AB radiation has been done quite recently \cite{ABrad1,ABrad2}.

In \cite{ookouchi}, based on the remarkable progress of the AB radiation, one of the authors studied cosmological constraints arising from the Big Bang Nucleosynthesis and the diffuse $\gamma$-ray background. Especially, in string theory such constraints are viable, and some of parameter spaces are excluded in some compactification scenarios.  In this paper, based on the study, we would like to go a step further toward an application to supersymmetric (SUSY) model building. One of the striking features of the SUSY model building is the existence of stabilized supersymmetric particles. Throughout this paper, we simply assume that an AB string associated with R-symmetry\footnote{A cosmic string associated with R-symmetry has been studied in \cite{Rstring}. Rich physical aspects such as instability of metastable vacua induced by the string and cosmological constraints coming from R-axion radiated by the string have been discussed.} is formed at the early stage of the universe and that the lightest supersymmetric particle carries a charge of the corresponding discrete symmetry.

The organization of this paper is as follows. In section 2, we briefly review the universal effective Lagrangian of $ \mathbb{Z}_p$ gauge theory, and show a relationship between AB strings/particles and the discrete gauge theory. Then, we summarize the results of calculations on the power of the AB radiation shown in \cite{ABrad1,ABrad2,ookouchi}. In section 3, we impose a cosmological constraint arising from the observed dark matter density. Section 4 is devoted to conclusions and comments on an application to string theories. In appendix A, we exhibit explicit calculations of AB radiation of massive spin $3/2$ particles. In appendix B, we briefly summarize the loop number density for cosmic strings loosing the energy via particle and gravitational radiation.

\section{Review of Aharonov-Bohm radiation}

In this section, we first review the universal effective Lagrangian of $ \mathbb{Z}_p$ discrete gauge theory and discuss Aharonov-Bohm (AB) particles/strings associated with the symmetry along the lines of \cite{BanksSeiberg}. The effective Lagrangian is described by BF coupling (or St\"uckelberg coupling),
\begin{equation}
p \int_{4D } {B}_2\wedge dA~,
\end{equation}
where $A$ is the massive gauge field one-form and $B_2$ is the Kalb-Ramond two-form field. The gauge transformation for each field is 
\begin{eqnarray}
&A \to A +d \lambda,\quad & \phi \to \phi +p \lambda , \nonumber \\
&{B}_2\to {B}_2 +d \Lambda ,\quad  & V \to V+p\Lambda , \label{gaugetr}
\end{eqnarray}
where $\Lambda$ is a one-form. The dual one-form gauge field $V$ transforms non-linearly, indicating the breaking of continuous $U(1)$ symmetry. Also, $\phi$ is the dual field of the Kalb-Ramond field $B_2$.

Following the arguments shown in \cite{BanksSeiberg}, let us review an interaction between AB strings and AB particles in $ \mathbb{Z}_p$ gauge theory. An AB particle is a particle-like object coupling electrically to the massive gauge field $A$. By exploiting a closed world-line or an infinite length of world-line (we denote $\sigma_1$), the AB particle can be written as a line operator,  
\begin{equation}
{\cal O}_{ particle}\sim {\rm exp}\left( {i\int_{\sigma_1}A} \right),
\end{equation}
where we assumed the minimum charge\footnote{Hereafter, we assume that AB particles and AB strings carry the minimum charges in the fundamental unit.}. On the other hand, an AB string is a string-like object coupling to $B_2$ electrically. In the same way, an operator of the AB string can be represented as a surface operator,
\begin{equation}
{\cal O}_{ string}\sim {\rm exp} \left( i \int_{\sigma_2} B_2   \right),
\end{equation}
where $\sigma_2$ is a closed surface or an infinitely large world-sheet. Note that $p$ AB strings annihilate with the junction operator $e^{-i \int_{L}V}$: Clearly, this junction operator is not invariant under the gauge transformation \eqref{gaugetr}. To compensate the non-invariance, one can add the world-sheet operator as $  {\rm exp} [ {-i \int_{L}V +ip \int_{C}B_2} ]$, where $\partial C=L$. This imply that  $p$ world-sheets corresponding to AB strings annihilate at the boundary $\partial C$ with the junction operator, indicating $p$ periodicity of $ \mathbb{Z}_p$ theory. To see the topological interaction between the AB string and the AB particle, let us put the AB string with the minimum charge in the space-time. The action becomes
\begin{equation}
p \int_{4D } B_2\wedge dA  + \int_{\sigma_2 } B_2.
\end{equation}
Consider the holonomy picked up by the AB particle circling around the AB string,
\begin{equation}
{ hol}(c)\equiv \exp \left( \int_{c} A \right)=\exp \left( \int_{S} F\right) =\exp \left( {2\pi i \over p} \right) \equiv \exp( i \phi),
\end{equation}
where $\partial S =c$. We used the equation of motion for $B_2$ in the third equality. We refer to the total magnetic flux in the AB string as $\phi$. When $p>1$, by means of the topological term in the Lagrangian, a non-trivial gauge potential is generated around the AB string, which gives rise to the Aharonov-Bohm effect. The interaction between the AB particle and the massive gauge field can be simply understood as $\int_{4D} A \wedge *_4  {\cal J.}$\footnote{Here, $*_4$ is the Hodge dual in four dimensional space-time.}

Now we are ready to review the total power of radiated particles from Aharonov-Bohm strings studied initially in \cite{ookouchi}. Basically, exploiting the analysis of \cite{ABrad1,ABrad2}, one can evaluate the radiation power since in the present situation, radiated particles are massive but much lighter than the scale of the string tension. According to the results shown in \cite{ABrad1,ABrad2}, the dominant radiation of massive particle comes from cusps (or kinks) on loops. Hence, we simple apply the formulae in \cite{ABrad1,ABrad2} for a cuspy loop to the current analysis. Also, as in the previous work \cite{ookouchi}, since we are interested in order estimation of cosmological constraints arising from the AB radiation, we will not carefully treat order one coefficients of the formulae.

From the equation of motion for the Kalb-Ramond field $B_2$, we obtain the gauge potential around the AB string,
\begin{equation}
p\, \partial_{\nu }A_{\mu} = \tilde{J}_{\mu \nu}. \label{topological}
\end{equation}
Here, we ignored the kinetic term which is irrelevant in our assumption of the string tension. In the wire-approximation, the dual of the string current is written as
\begin{equation}
\tilde{J}_{\mu \nu}= \epsilon_{\mu \nu \alpha \beta} \int d \tau d \sigma (\dot{X}^{\alpha}{X^{\prime}}^{\beta}-\dot{X}^{\beta} {X^{\prime}}^{\alpha} ) \delta^{(4)} (x-X(\sigma , \tau)). \label{dualstringcurrent}
\end{equation}
$\sigma, \tau$ are the world-sheet coordinates of the string. In momentum space, the solution of \eqref{topological} is written as follows: 
\begin{equation}
A_{\mu}={1\over p} \epsilon_{\mu \nu \alpha \beta}  {k^{\nu} \over (k_{\lambda}k^{\lambda} ) } J^{\alpha \beta}. \label{ABsolution}
\end{equation}
This gauge potential is the same as the one for the solenoid shown in \cite{AlWil, ABrad1,ABrad2}. Hence, we can proceed along the lines of \cite{AlWil, ABrad1,ABrad2} to obtain the emission rate of AB particles from a single cusp. Below, we will mainly discuss emissions of spin $1/2$ fermions, so the current coupling to the gauge potential is given by ${\cal J}_{\mu}=\sum_a \bar{\psi}_a \gamma_{\mu} \psi_a $. In this case, according to \cite{ABrad1,ABrad2},  the radiation power from a single cusp is 
\begin{equation}
P_{ AB}= \sum_{k=1}^{K_{max}} P_{AB}^{(k)}  \simeq   \sum_{k=1}^{K_{max}} \Gamma_{ AB} { \phi^2 \over  L^2} = \Gamma_{ AB} { \phi^2 \over  L^2}K_{max}, 
\end{equation}
where $\mu$ is the tension of the AB string. $\Gamma_{ AB}$ is the numerical coefficient depending on dynamics of strings and we roughly estimate as ${\cal O}(10^{-5\mbox{-}-2})$ \cite{NewVacha1}. $k$ represents a mode included in the Fourier expansion of a string configuration near a cusp. It is interesting that the power for each mode, $P_{AB}^{(k)}$, does not depend on $k$, so the total power is highly sensitive to the ultraviolet cut-off $K_{max}$. Note that for loops with the size $L$, the typical energy scale of radiated particles ${\cal O}(k/L)$ is much larger than the scale from na\"ive dimensional analysis ${\cal O}(1/L)$ because cusps include high frequency modes $K_{ max}\gg 1$. Remarkably, the maximum number of modes is much larger than the one from na\"ive expectation \cite{NewVacha1,Vach}. The breakdown scale of the wire-approximation gives us the na\"ive maximum number $N_{naive}\sim  \sqrt{\mu}L$. However, as is pointed out in \cite{NewVacha1,Vach}, the speed near a cusp reaches the speed of light, so the boost factor enhances the maximum number of modes as\footnote{This power can be understood in the following way. From na\"ive estimation, the maximum momentum of radiated particles is ${\cal O}(\sqrt{\mu})$. However, by the boost factor, it gets the extra factor $(\sqrt{\mu}L)^{1/2}$ and becomes ${\cal O}(\sqrt{\mu}(\sqrt{\mu}L)^{1/2})$. Using the definition of $N_{naive}$, one find $N_{naive}^{3/2}/L$. } $K_{ max}\sim N_{naive}^{3/2}$. Therefore, the power of the AB radiation from the single cusp is given by
\begin{equation}
P_{ AB} \simeq  \Gamma_{ AB} { \phi^2 \over  L^2}(\sqrt{\mu}L)^{3/2}=\Gamma_{ AB} { \phi^2 \mu^{3/4}\over  \sqrt{L}}.\label{powerABrad}
\end{equation}
Although \eqref{powerABrad} is the formula for massless radiation, by recalling that radiated particles are lighter than the string scale, we find that it can be reliably applicable to the present massive radiation. As for the other emissions such as radiation from a kink or a kink-kink collision, the loop size dependences of the power are different 
\begin{equation}
P_{ AB}{(p)} \simeq \Gamma_{ AB}^{(p)} { \phi^2 \mu   \over (\sqrt{\mu} L)^{p}},\label{VariousRad}
\end{equation}
where $p=4/3$ for a kink and $p=1$ for a kink-kink collision \cite{NewVacha1}.

One more relevant radiation which is common in comic strings is gravitational wave radiation. Since graviton is massless, the main radiation comes from oscillations of loops. The power of gravitational radiation is given by (see \cite{VS} for example), 
\begin{equation}
P_{ GW} = \Gamma_{ GW} \,G \mu^2.  \label{GWpower}
\end{equation}
$G$ is the Newton constant\footnote{For later reference, $G=1/M_{pl}^2$. } and $\Gamma_{ GW} = {\cal O}(10)$ is a constant depending on dynamics of string network. From \eqref{powerABrad} and \eqref{GWpower}, we find that gravitational radiation is dominant for large size of loops. The critical length is given by 
\begin{equation}
(L_{ crit})^p= {\Gamma_{ AB}^{(p)} \phi^2 \over \Gamma_{ GW}}  (G\mu)^{-1-{p\over 2}} G^{p\over 2}.
\end{equation}

To calculate the total power of AB radiation from all loops, we need to know the number density of loops. Here, we assume that string network reaches the scaling regime as in the standard cosmic strings \cite{VS}. As we review in appendix B,  the number density of loops with the size between $L $ and $L+d L$ is given by, 
\begin{equation}
 n_{L} = {1\over p_{st}} \times \left \{ 
 \begin{array}{l} 
 n_{GW}(t, L) \qquad  ({\rm GW\, rad \, dominant}) \\ 
  n_{AB}(t, L)\, \qquad  ({\rm AB\, rad \, dominant})
 \end{array} 
 \right. \label{number2}
\end{equation} 
where $n_{AB}$ and $n_{GW}$ are defined by \eqref{nGW} and \eqref{nAB}. We will soon write down appropriate expressions for our calculations below.  In \eqref{number2} we included the reconnection probability $p_{st}$ of cosmic strings that is relevant in applications to string theories where the reconnection probability is smaller than unity \cite{CSSreview}. By using the number density of loops, we calculate the total power of radiation from all loops,
\begin{eqnarray}
\dot{\rho}(t)&=&B_{ r} m_{LSP}\, \dot{n}(t)=B_{ r}  m_{LSP} \int_{L_{min}}^{L_{ max}} dL \, n_L \sum_{k=1}^{K_{max}} {L\over k}  P^{(k)}_{ AB}, \nonumber \\
&=& B_{ r}  m_{LSP} \int_{L_{ min}}^{L_{ max}} dL \, n_L {\Gamma_{ AB}\phi^2 \over L}\cdot \beta ,\qquad {\rm where}\quad \beta\equiv \sum_{k}^{K_{max}} {1\over k}.\label{rhodot}
\end{eqnarray} 
According to the recent computer simulations of string networks \cite{simulation1, simulation2}, the largest size of loops is given by $L_{ max}\simeq \alpha t$ with $\alpha \sim 0.1$. $\dot{n}(t)$ is the number density of radiated particles per time and $k/L$ is a typical momentum of radiated particles from the $k$-th Fourier mode\footnote{This is the mean comoving momentum. As discussed in \cite{Kawasaki}, radiation of Nambu-Goldstone bosons from global strings, the comoving momentum is estimated as ${\cal O}(1/t)\simeq {\cal O}(  1/L)$. In the present situation, the dominant radiation comes from high frequency modes included in cusps, so we should account for the extra mode factor $k$. }. It is worthy emphasizing that from the formula, $\beta\equiv \sum_{k}^{K_{max}}{1\over k}\simeq \log K_{max}$, we find that contribution from each mode to the number density per unit time is the same order and does not strongly depends on $K_{max}$. We think that the dominant contribution to the dark matter density comes from momentum modes comparable to the mass of the dark matter because higher modes tend to be converted to the production of other particles. So, in order to incorporate the energy loss of the high frequency modes, we introduced the branching ratio $B_r<1$ as a tunable parameter and will estimate cosmological constraints by changing the parameter.  In this case, we can simply obtain the energy density per unit time by multiplying the mass of the dark matter. This assumption is consistent with the computer simulation for axion production from a global string \cite{Kawasaki}. Moreover, we would like to comment on the reason why the lower bound of the integral is given by $L_{min}$ which is defined by the following expression. This length is determined by the condition for loops with shorter life-time than one Hubble time, $P^{-1}_{GW/AB}\mu L <t$,
\begin{equation}
L_{min}={\rm max}[  L_{AB} , L_{GW}   ]\equiv {\rm max} \left[({\Gamma_{AB}^{(p)}\phi^2 t/ \mu^{p/2}})^{1/(1+p)} ,\Gamma_{GW} G\mu t\right] ,\label{minisize}
\end{equation}
where ${\rm max}[A,B]= A$ or $B$ whichever is larger. By using the explicit expressions of the loop number density \eqref{number2} and the power of radiation \eqref{powerABrad} and \eqref{GWpower}, one can easily find that $L$ dependence of the integral for shorter loops,
\begin{equation}
 \int_{L_*}^{L_{min}} dL \, n_L  L P^{(k)}_{AB} \sim \left \{ 
 \begin{array}{l} 
 \Big[ \log L \Big]^{L_{min}}_{L_*} \ (L_{AB}<L_{GW}) \\ 
 \Big[ L^{1/2}  \Big]^{L_{min}}_{L_*} \ (L_{GW}<L_{AB})
 \end{array} 
 \right. ,\label{mushi}
\end{equation} 
where $L_*$ is the IR cut-off that is expected to be of order ${\cal O}(\mu^{-1/2})$. Since \eqref{mushi} is a monotonically increasing function of $L$, we conclude that contributions from loops with short size $L\ll L_{min}$ become subdominant. For the aim of rought order estimation, we can take the lower bound of the integral in \eqref{rhodot} $L_{min}$. In the interval $[L_{min}, L_{max}]$, the loop number density \eqref{nGW} and \eqref{nAB} can be simplified as follows:
\begin{equation}
 \left \{ 
 \begin{array}{l} 
 n_{GW}(t<t_{eq},L) = \frac{\zeta \sqrt{\alpha}  t^{-3/2}}{ L^{5/2}} \Theta \left( 1-\frac{L}{\alpha t}\right) \\ 
 n_{GW}(t_{eq}<t,L) = \frac{\zeta \sqrt{\alpha}  t_{eq}^{1/2}t^{-2}}{ L^{5/2}} \Theta(1-\frac{ L }{\alpha t_{eq}})  +  {1 \over t^2} {\zeta_m \sqrt{\alpha} \over L^2}
 \end{array} 
 \right. \label{nGW},
\end{equation} 
\begin{equation}
 \left \{ 
 \begin{array}{l} 
 n_{AB}(t<t_{eq},L) = \frac{\zeta \sqrt{\alpha}  t^{-3/2}}{ L^{5/2}} \Theta \left(1-\frac{L}{\alpha t }\right)  \\ 
 n_{AB}(t_{eq}<t,L) = \frac{\zeta \sqrt{\alpha} t_{eq}^{1/2} t^{-2}}{ L^{5/2}} \Theta \left(1-\frac{  L}{ \alpha t_{eq} }\right)+ {1 \over t^2} {\zeta_m \sqrt{\alpha}\over L^{2}}
 \end{array} 
 \right.  \label{nAB}.
\end{equation} 
Finally for later reference, we show the critical time when $L_{GW}$ equals to $L_{AB}$,  
\begin{equation}
t_{crit}\equiv 5.4\times 10^{-44} {\rm [s]} {(\Gamma_{AB}^{(p)} \phi^2)^{1/p} \over (\Gamma_{GW}  G\mu)^{1+{1\over p}}} (G\mu)^{-1/2}. \label{crittime}
\end{equation}

\section{Cosmological constraint}

Now we are ready to estimate cosmological constraints. In this section, we discuss constraints coming from the observed dark matter density and big bang nucleosynthesis. 

\subsection{Dark matter constraint}

We evaluate a cosmological constraint by using the formulae for radiation from cusps $p=1/2$, which is the dominant radiation, shown in the previous section. From the cosmological observation of dark matter density \cite{WMAP7},
\begin{equation}
{\rho_c\over s_0}  \simeq 4.7\times 10^{-10}  {\rm [GeV]} \left({\Omega_m h^2\over 0.13} \right), \label{WMAPdata}
\end{equation}
we obtain a constraint on parameters of model building. $\rho_c$ and $s_0$ are the dark matter density and the entropy density in the present age. Here without specifying the origin\footnote{For an example, see Appendix in \cite{ookouchi}. R-strings formed in compactifying Type IIA string theory were discussed.}, we simply assume that a discrete gauged R-symmetry exists and that an Aharonov-Bohm R-string associated with the symmetry is created in the early stage of the universe. We denote the time of string creation $t=t_{string}$. Also, we suppose that the lightest supersymmetric particle (LSP) is a stable fermion such as neutralino. In this case, radiation of LSP from the Aharonov-Bohm string offers non-thermal production of dark matter\footnote{See \cite{DarkHind} and references therein for recent studies on dark matter productions from cosmic strings.}. Since AB radiation is a pair production of particle and anti-particle, we should be careful about the freeze-out time (or temperature) of LSP. Large amount of extra production of the particles/anti-particles enhances annihilation process, and reduces the total abundance of the particles. So the starting time of accumulation of LSP should be given by 
\begin{equation}
t_i\equiv {\rm max}[t_{string}, t_{freeze}],
\end{equation}
where $t_{freeze}$ is the freeze-out time after taking into account the non-thermal production of dark matter arising from the AB radiation. Here, we do not count the radiated particles before the freeze-out time. This is because for a conservative estimation of cosmological constraint. Before freezing-out, LSP can annihilate and the energy can be converted to other particles. However, this process highly depends on strengh of interactions in models. So for precise estimation, we need to specify the interactions of LSP with standard model particles. In this paper, since we would like to find a generic constraint related to AB radiation, so we simply ignore the remnant of LSP radiated before the freeze-out time, and show a conservative cosmological constraint. Also, in supersymmetric model building, usually, the hidden sector is not thermalized to avoid overproduction of gravitinos or stable supersymmetric particles. In this case, LSP radiated before freeze-out time is not in thermal bath, which makes estimation of the remnant LSP involved. So, below, for the sake of simplicity, we assume the starting time of accumlation of LSP is the freeze-out time.

\begin{figure}[htbp]
\begin{center}
 \includegraphics[width=.4\linewidth]{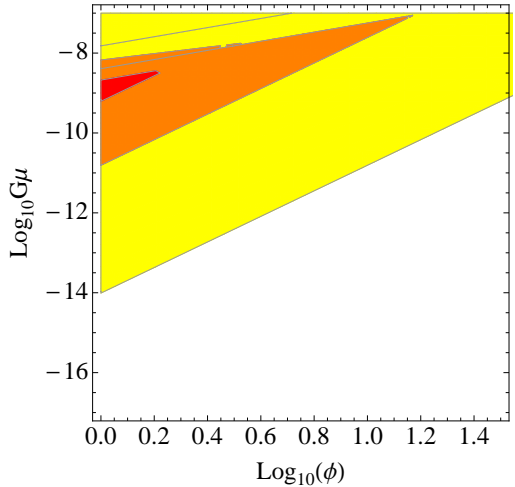}
\vspace{-.1cm}
\caption{\sl  Constraints from the observed dark matter density. We chose $\Gamma_{GW}=50$,  $\Gamma_{AB}=10^{-4}$, $c=c^{\prime}$ and $c \zeta\beta B_{ r}{m_{LSP}}/M_{pl}=10^{-16}$. The red, orange and yellow regions are excluded by the conditions for $t_i=10^{-26}{\rm [s]}$, $10^{-27}{\rm [s]}$ and $10^{-29}{\rm [s]}$, respectively. The white regions is allowed.  }
\label{FIG1}
\end{center}
\end{figure}

Following the strategy used in \cite{BL}, we consider time integral of $\dot{\rho}/s$ and imposes the constraint \eqref{WMAPdata}. It is useful to divide it into three parts,
\begin{equation}
I\equiv \int_{t_i}^{t_0} {\dot{\rho}\over s} \, dt = \int^{t_{crit}}_{t_i} {\dot{\rho}\over s} \, dt +\int_{t_{crit}}^{t_{eq}} {\dot{\rho}\over s}\, dt +\int_{t_{eq}}^{t_0} {\dot{\rho}\over s} \, dt \  < \, {\rho_c\over s_0}. \label{threerad}
\end{equation}
The entropy density in the radiation/matter dominated era is given by
\begin{equation}
 s(t) \simeq \left \{ 
 \begin{array}{l} 
 0.13(G t^2)^{-3/4} \qquad   {\rm for}\quad {\rm radiation\ dominant} \\ 
 0.07(G T_{eq})^{-1} t^{-2}  \quad  {\rm for}\quad  {\rm matter\ dominant} 
 \end{array} 
 \right. .
\end{equation} 
Plugging \eqref{number2} and \eqref{minisize} into \eqref{threerad}, we explicitly calculate the total radiation power. Let us start with the first integral in \eqref{threerad},
\begin{equation}
I\simeq \int^{t_{crit}}_{t_i} {\dot{\rho}\over s}\,  dt \, \simeq \, c\times 10^{-9} {\rm [GeV]}B_{ r} {m_{LSP}\over M_{pl}}  {\zeta \beta \sqrt{\alpha}\over (\Gamma_{AB}\phi^2)^{2/3}} (G\mu)^{5/12} \left( {1 {\rm [s]}\over t_i}\right)^{2/3}   \quad  t_i<t_{ crit}, \label{result1}
\end{equation}
where $\beta\equiv \sum_{k}^{K_{max}}{1\over k}\simeq \log K_{max}$. $c$ is order one numerical coefficient. Although $\beta$ is a function of $G\mu$ and $L$, we evaluate it as a parameter of order ${\cal O}(10^{1\mbox{-}2})$ since their dependences are $\log$. This is enough for order estimation of constraints. The dominant contribution comes from the lower bound of the integral. In fact, one can easily check that this is the dominant for all three integrals in \eqref{threerad} for the parameter region $t_i< t_{ crit}$. From \eqref{crittime}, one finds that in the region of large $G\mu$, this condition, $t_i< t_{ crit}$, does not tend to be satisfied. Actually, as we will exhibit in the figure \ref{FIG1}, there is parameter region where this condition is not satisfied. In this case, the first term of \eqref{threerad} does not exist and the lower bound of the second integral becomes $t_{i}$ instead of $t_{crit}$. Again, the dominant contribution comes from the lower bound of the second integral,
\begin{equation}
I\simeq \int_{t_{i}}^{t_{eq}} {\dot{\rho}\over s}\,  dt\,  \simeq\, c^{\prime} \times 10^{-46} {\rm [GeV]} B_{ r}{m_{LSP}\over M_{pl}} \zeta \beta \sqrt{\alpha} {\Gamma_{AB}\phi^2 \over \Gamma_{GW}^{5/2}}(G\mu )^{-5/2} \left({1 {\rm [s]} \over t_i} \right)^{3/2} \quad {\rm for} \quad  t_{ crit}<t_i ,\label{result2}
\end{equation}
where $c^{\prime}$ is order one numerical coefficient. From \eqref{result1} and \eqref{result2}, we find that $G\mu$ dependences are different from each other. In the large $G\mu$ region, as we decrease the scale of the tension, the constraint becomes sever since emissions of particles are enhanced. However, at some point, the dominant contribution switches to \eqref{result1} which reduces particle emission as $G\mu$ decreases. So the excluded region forms a kind of strip shape. As an illustration, we exhibit the excluded regions for the cases $t_i=10^{-29}$, $10^{-27}$ and $10^{-26}$[s] with the parameters $\Gamma_{GW}=50$,  $\Gamma_{AB}=10^{-4}$, $c=c^{\prime}$ and $c \zeta\beta {B_r m_{LSP}}/M_{pl}=10^{-16}$. It is interesting that constraints are weak but non-negligible to exclude parameter space of the string tension. One way to avoid such exclusions is to engineer a phenomenological model in which the freeze-out time is late enough.

\subsection{Constraint from big bang nucleosynthesis}

As we have shown, as long as the branching ratio $B_r$ is not so small, some parameter regions, though they are very narrow windows, are constrained by the observation of the dark matter density. On the other hand, since the radiated particles are relativistic, it is likely that fraction of radiated energy is converted into production of the standard model particles. In this case, we can, in turn, impose other constraints arising from the big bang nucleosynthesis or cosmic microwave background. Since these constraints becomes sever when the branching ratio of dark matter radiation is small, they can be regarded as complementary conditions\footnote{ We would like to thank the referee for pointing us on this possibility.}. There are two kinds of radiation processes of the standard model particles. One is secondary radiation from supersymmetric particles emitted firstly from cosmic strings. The other is direct radiation of the particles from the strings. The interactions of the standard model particles with LSP or cosmic strings are highly model dependent, so we cannot know them without specifying the model. Here, we simply assume the existence of the interactions and control the strength of the interactions by branching ratio $B_r^{SM}$. After some numerical calculation, we find that the former radiation does not yield viable constraint, so we discuss only the latter radiation below. It is worthy emphasizing that BBN constraint, which will be discussed below, requires injected total energy when the light elements are forming. Therefore, contributions from the high frequency modes play important role. BBN constraint without the boost factor in $K_{max}$ was studied in \cite{ookouchi}, so we here focus on the case with the boost factor.  
 
\begin{figure}[htbp]
\begin{center}
 \includegraphics[width=.4\linewidth]{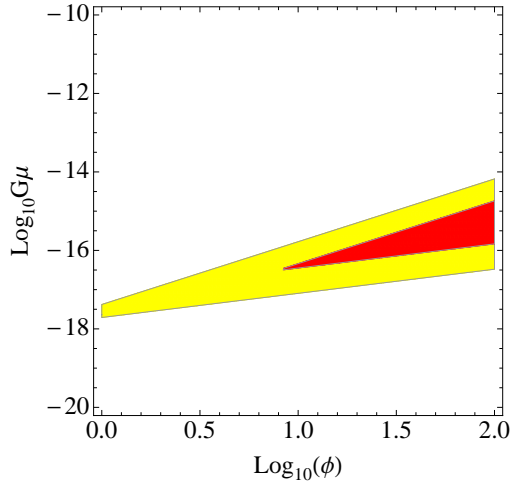} 
 \vspace{-.1cm}
\caption{\sl Constraints from leptonic radiation at $t=10^7${\rm [s]}. We took $\Gamma_{AB}=10^{-3}$, $\Gamma=50$, ${B_r^{SM}\tilde{c }\xi \sqrt{\alpha}\over 2p_{st}}=2,10$ for red and yellow.}
\label{lepton}
\end{center}
\end{figure}

We can proceed the study along the lines of \cite{Dufaux}. If muons or photons are radiated, the light elements formed by BBN will be destroyed. Here, for the sake of simplicity, we show constraint coming from radiation of leptons. The upper bound of injected energy is given by \cite{BBN},
\begin{align}
\label{BBNconstraint}
\frac{\rho}{s}<\left\{
\begin{array}{ll}
10^{-8}\left({t\over 10^{4} {\rm [s]} } \right)^{-2}\,{\rm [GeV]} & {\rm for} \quad 10^4 {\rm [s]}  \le t < 10^{7} {\rm [s]}, \\
10^{-14}\, {\rm [GeV]} &{\rm for}  \quad 10^7{\rm [s]}  \le t \le 10^{13} {\rm [s]}.\\ 
\end{array}\right.  
\end{align}
It is convenient to use the energy injected energy in one Hubble time, $\rho_{SM}= t \dot{\rho}$, and divide it by the entropy density, $s=(g_*/45\pi)^{1/4}(M_{ pl}/t)^{3/2}/4$, 
\begin{equation}
{\rho_{SM} \over s}\simeq 1.1\times 10^{-45}     \left({10.75 \over g_*} \right)^{1/ 4}\left(  {t \over 1 {\rm [s]}}  \right)^{5/2}\left(  {\hbar \over 6.5\times 10^{-16} {\rm [eV\cdot s]}}  \right)^{3/ 2} {\rm [GeV]} \left[ {1 {\rm [s]^4}\over M_{ pl}} \dot{\rho} \right],
\end{equation}
where $M_{pl}=G^{-1/2}\simeq1.2\times 10^{19}$ [GeV]. Using \eqref{powerABrad} and \eqref{number2}, we can estimate the energy as follows:
\begin{eqnarray}
{1 {\rm [s]}^4 \over M_{ pl}}   \dot{\rho}&=&{1 {\rm [s]}^4 \over M_{ pl}}B_r^{SM}  \int^{L_{ max}}_{L_{ min}} d L\, n_L P_{ AB} = {1 {\rm [s]}^4 \over M_{ pl}}{\tilde{c} \phi^2 B_r^{SM}  \xi \sqrt{\alpha} \over 2p_{st} t^{3/2}}\mu^{3/4}  \Gamma_{ AB} L_{ min}^{-2}+\cdots ,
\end{eqnarray}
where we denoted ${\cal O}(1)$ coefficient $\tilde{c}$. To impose the constraint, let us write down $\rho_{SM}/s$ first, 
\begin{eqnarray}
{\rho_{SM} \over s} \simeq \left\{
\begin{array}{ll}
\tilde{c}\times 10^{5} \left(  { 1 {\rm [s]}\over t}  \right)^{{1/ 3}} \left(  {10.75 \over g_*}  \right)^{1/ 4} {B_r^{SM}\xi \sqrt{\alpha}   \over 2 p_{st} } \phi^{-2/3}    \Gamma_ {AB}^{-1/3} (G\mu)^{13/12} \ {\rm [GeV]}   &  \ \text{for}\  L_{ min}=L_{ AB}  \\ 
\tilde{c} \times 10^{-24} \left(  {1 {\rm [s]}\over t}  \right)\left(  {10.75 \over g_*}  \right)^{1/ 4}  {B_r^{SM}  \xi \sqrt{\alpha}   \over 2 p_{st} }  ( \phi^{2} \Gamma_{ AB}){1\over \Gamma_{GW}^2}  (G\mu)^{-5/4}\  {\rm [GeV]}   &  \ \text{for}\ L_{ min}=L_{ GW}. \label{result}
 \end{array}\right. 
\end{eqnarray}
By choosing appropriate parameters, $\Gamma_{AB}=10^{-3}$, $\Gamma=50$, ${B_r^{SM}\tilde{c}\xi \sqrt{\alpha}\over 2p_{st}}=2,10$ and imposing \eqref{BBNconstraint}, we find that there are small but non-negligible windows. The most sever condition comes from the time $t=10^{7}$[s]. See figure \ref{lepton}. It is remarkable that contrary to the case without the boosted factor \cite{ookouchi}, there exist nonzero excluded regions without assuming small reconnection probability. This is because the boost factor at the high frequency modes enhances the injected energy. However, as one can see from the parameter choice, ${B_r^{SM}\tilde{c }\xi \sqrt{\alpha}\over 2p_{st}}=2,10$, such constraints are still very weak when the branching ratio is not so large.

\subsection{Gravitino radiation}

Finally, we would like to study gravitino radiation from an Aharonov-Bohm R-string. In a large class of supersymmetric models, gravitino is the lightest supersymmetric particle (see \cite{KOO} for a recent review). Gravitino is a spin $3/2$ fermion and is described by the Rarita-Schwinger equation. In a gauged R-symmetric theory, the associated massive gauge field is graviphoton which couples to gravitino as follows, 
\begin{equation}
{\cal L}\sim -{1\over 2}\epsilon^{\mu \nu \rho \sigma} \bar{\psi}_{\mu}\gamma^5 \gamma_{\nu}\psi_{\sigma} A_{\rho}\equiv {\cal J}^{\rho} A_{\rho}. \label{CurrentDef}
\end{equation}
Again, since we assume that the mass of the gauge field is much larger than the gravitino mass, gravitino is regarded as the massless field. However, as shown in Appendix A, there is a subtlety in taking the massless limit. There is a discontinuity between the amplitude for the radiation of the exactly massless gravitino and the one for the massless limit of the massive gravitino. This discontinuity gives rise to a different numerical factor of the AB radiation. However, as far as order estimation of radiated power is concerned, this subtlety is not essential. So, for the sake of simplicity, we here discuss the exactly massless case. See Appendix A for the amplitude of massive radiation. In the massless case, the Rarita-Schwinger equation, 
\begin{equation}
\epsilon^{\mu \nu \rho \sigma} \gamma^5 \gamma_{\nu}D_{\rho}\psi_{\sigma}=0,
\end{equation}
is invariant under the gauge transformation $\psi_{\mu}\to \psi_{\mu}+D_{\mu}\eta$. By using the gauge fixing condition $\gamma^{\mu}\psi_{\mu}=0$ \cite{PF}, one can show that the Rarita-Schwinger equation is reduced to the Dirac equation,
\begin{equation}
\gamma^{\rho}D_{\rho} \psi_{\sigma} =0.
\end{equation}
Hence, except an irrelevant numerical factor, we simply apply our estimation shown above to the gravitino radiation. Thus, from the figure \ref{FIG1}, we find that scale of string tension is also constrained for a supersymmetric phenomenological model with gravitino LSP.

\section{Conclusions and comments \label{stringApp}}

In this paper, we extended the studies on cosmological constrains arising from the Aharonov-Bohm radiation \cite{ookouchi} to supersymmetric model building. A discrete R-symmetry can stabilize the lightest supersymmetric particle, and radiation of the particle from the string gives us a non-thermal production of dark matter. Combining the observed data for the dark matter density, we showed that the tension of the AB string or the freeze-out temperature of the AB particle is constrained. Also, we investigated radiation of standard model particles from AB strings. When radiated particles are relativistic, it is plausible to assume that fraction of the total radiated energy is converted into production of the particles. We find that there is a narrow excluded window from cosmological constraint arising from the big bang nucleosynthesis. Remarkably, contrary to the previous work \cite{ookouchi}, such constraints are non-negligible even within the field theory. In the context of string theories, cosmological constraints become sever owing to the small reconnection probability. Below, we would like to comment on a potential application of our analysis to string theories. 

\begin{figure}[htbp]
\begin{center}
 \includegraphics[width=.4\linewidth]{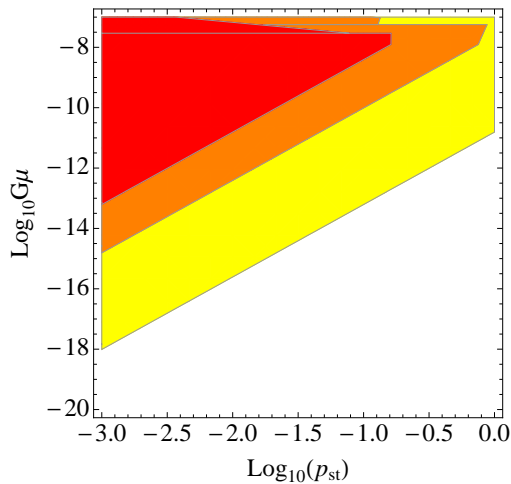}
\vspace{-.1cm}
\caption{\sl  Constraints from dark mater density. We took $\phi=10$, $\Gamma_{GW}=50$, $\Gamma_{AB}=10^{-4}$, $c=c^{\prime}$ and $c \zeta \beta {m_{LSP}}/M_{pl}=10^{-16}$. The red, orange and yellow regions are excluded by the condition \eqref{threerad} for $t_i=10^{-26}{\rm [s]}$, $10^{-27}{\rm [s]}$ and $10^{-29}{\rm [s]}$, respectively. 
 }
\label{FIG2}
\end{center}
\end{figure}

In string theories\footnote{See \cite{Uranga} for recent studies of discrete gauge symmetries in string theories.}, we have two clear differences. One is the Newton constant of four dimensional theory, $G=l_{pl}^2$, which is written in terms of the string length $l_s$ and the size of compactification scale $l_c$ as $l_s^8 =G l_c^6$. Since we assume $l_c> l_s$, we are left with the relation, $l_{pl} < l_s < l_c$. The other new feature is the reconnection probability $p_{st}$. As is mentioned below \eqref{number2}, the existence of the extra dimension makes the reconnection probability smaller than unity. Smaller probability increases the production rate of AB particles since the loop number density increases with $p_{st}^{-1}$. As an illustration, we exhibit the constraint \eqref{threerad} in the figure \ref{FIG2}. We choose the parameters $\phi=10$, $\Gamma_{GW}=50$, $\Gamma_{AB}=10^{-4}$, $c=c^{\prime}$ and $c \zeta \beta {m_{LSP}}/M_{pl}=10^{-16}$.

As a concrete example, let us consider KKLT scenario \cite{KKLT,KKLMMT} in which the warped conifold \cite{KS} is used for Type IIB string compactification. In the context of gauge/gravity duality, a cone-type throat is identified with a realization of super-conformal symmetry, which includes R-symmetry, in four dimensional field theory. At the tip of the conifold, the continuous $U(1)_R$ symmetry is broken to $ \mathbb{Z}_2$ symmetry by the deformation of the conical singularity. In KKLT scenario, since the total manifold is compact, the $ \mathbb{Z}_2$ symmetry is gauge symmetry. A natural candidate for the Aharonov-Bohm string associated with the gauged $ \mathbb{Z}_2$ theory is a D1 string or a its bound state with an F1 string, namely a $(p,q)$-string \cite{Rob}\footnote{See \cite{Polchinski} for earlier studies of cosmic superstring and \cite{CSSreview} for reviews.}. When the numbers $(p, q)$ and $g_s$ are of order one, the tension of the $(p,q)$ string is roughly given by  $\mu_{p,q}\sim {f/ l_s^2} $ where we included the warp factor $f\ll 1$.

From the figure \ref{FIG2}, suppose that we have the excluded region $10^{-18}<G\mu_{p,q}< 10^{-8}$. This condition is translated into that of the warp factor as follows:
\begin{equation}
10^{-18} \left( {l_c\over l_s} \right)^6 <f <  10^{-8} \left( {l_c\over l_s} \right)^6.
\end{equation}
To be more concrete, we further assume that the compactification scale is of order $l_c=100l_s$. In this case, the range of the warp factor, $10^{-6}  <f < 10^{4}$, is excluded.  
It is interesting that natural parameter region can be excluded. Thus, to engineer realistic model in the KKLT scenario, we should be careful about the freeze-out temperature of LSP or the order of the wrap factor. It would be interesting to explore further in string model building by specifying the model explicitly. Moreover, cosmological constraints arising from emissions of standard model particles such as Higgs bosons and Z-bosons were discussed in \cite{NewVacha1,NewVacha2,Hind} recently. Applying these approaches to cosmic superstrings would give us stringent constraints for sting compactifications. We will leave them for the future publication.

\section*{Acknowledgement}

The authors are grateful to Masahiro Ibe, Ryuichiro Kitano and Ken-ichi Okumura  for useful comments and discussions. The authors would like to thank the Yukawa Institute for Theoretical Physics at Kyoto University for hospitality during the workshop YITP-W-14-4 ``Strings and Fields'', where part of this work was carried out. This work is supported by Grant-in-Aid for Scientific Research from the Ministry of Education, Culture, Sports, Science and Technology, Japan (No. 25800144 and No. 25105011).

\appendix

\setcounter{equation}{0}
\renewcommand{\theequation}{A.\arabic{equation}}
\section*{Appendix A\, Radiation of massive gravitinos}

In this appendix, for completeness, we calculate the amplitude for AB radiation of massive spin $3/2$ particles along the lines of \cite{ABrad1,ABrad2}. By adding the mass term for spin $3/2$ fermion, the Lagrangian becomes
\begin{equation}
{\cal L}=-\frac{1}{2}\epsilon^{\mu \nu \rho \sigma} \Bar{\psi}_{\mu}(x) \gamma^{5}\gamma_{\nu} D_{\rho} \psi_{\sigma}(x) -\frac{1}{4} m \Bar{\psi}_{\mu}(x) [\gamma^{\mu},\gamma^{\nu}]\psi_{\nu}(x)  ,
\label{Lag}
\end{equation}
where $D_{\rho}=\partial_{\rho} -i e A_{\rho}$ the covariant derivative\footnote{We use the notation $\{\gamma^{\mu}, \gamma^{\nu}\}=2g^{\mu \nu}$. The metric $g_{\mu \nu}$ is ${\rm diag}(1,-1,-1,-1)$ in our convention. }. Below, we calculate the tree level amplitude of particle emissions from an AB string, so it would be useful to discuss first the free massive spin $3/2$ field. The Rarita-Schwinger equation for the free field can be simplified by using the following relation,
\begin{equation}
\gamma_{\mu} \epsilon^{\mu \nu \rho \sigma}=-i\gamma^{5}(\gamma^{\nu} \gamma^{\rho} \gamma^{\sigma}-g^{\rho \sigma} \gamma^{\nu}+g^{\nu \sigma}\gamma^{\rho} - g^{\nu \rho} \gamma^{\sigma}).
\label{shiki1}
\end{equation}
The equation of motion is rewritten as 
\begin{equation}
i(\gamma^{\mu}\Slash{\partial}\gamma^{\nu}\psi_{\nu}-\gamma^{\mu}\partial^{\nu}\psi_{\nu}+\Slash{\partial}\psi^{\mu}-\gamma^{\nu}\partial^{\mu}\psi_{\nu})+m(\gamma^{\mu}\gamma^{\nu}\psi_{\nu}-\psi^{\mu})=0.
\label{Eom1}
\end{equation}
Acting $\partial_{\mu}$ and $\gamma_{\mu}$ on \eqref{Eom1}, respectively, we obtain
\begin{align}
\label{Eom2}
& m(\Slash{\partial}\gamma^{\nu}\psi_{\nu}-\gamma^{\nu}\Slash{\partial}\psi_{\nu})=0, \\
\label{Eom3}
& i(\Slash{\partial}\gamma^{\nu}\psi_{\nu}-\gamma^{\nu}\Slash{\partial}\psi_{\nu})+3m\gamma^{\nu}\psi_{\nu}=0.
\end{align} 
From these equations, it is easy to derive the following equations
\begin{equation}
  \left \{ 
 \begin{array}{l} 
\gamma^{\mu}\psi_{\mu}(x)=0 \\ 
(i\Slash{\partial}-m)\psi_{\mu}(x)=0\\
\partial^{\mu}\psi_{\mu}(x)=0
 \end{array} 
 \right. ~.
\end{equation} 
The first equation corresponds to the gauge fixing condition for the massless field shown in the main text. The second equation is the Dirac equation for the massive spinor field. In the same ways as the free Dirac spinor field, it is useful to expand a solution of these equations by positive and negative frequency modes 
\begin{equation}
\psi_{\mu}(x)=\left\{ \begin{array}{ll}
\psi_{\mu}^{(+)s}(p) e^{-ip\cdot x} \\
\psi_{\mu}^{(-)s}(p) e^{+ip\cdot x} \\
\end{array} \right. .
\label{kai}
\end{equation} 
Here, we impose the following normalization conditions on $\psi_{\mu}^{(+)s}$ and $\psi_{\mu}^{(-)s}$,
\begin{equation}
 \Bar{\psi}_{\mu}^{(+)r}(p)\psi^{(+)s\mu}(p)=-2m\delta^{r,s}, \qquad \Bar{\psi}_{\mu}^{(-)r}(p)\psi^{(-)s\mu}(p)=+2m \delta^{r,s}.
\label{kikakuka}
\end{equation}
With these relations, it is easy to see that the current with the spin polarization 
\begin{equation}
{\cal J}^{\rho}_{ss^{\prime}}\equiv g^{\mu\sigma}\Bar{\psi}_{\mu}^{(+)s}(k)\gamma^{\rho}\psi_{\sigma}^{(-)s^{\prime}}(k^{\prime}),
\end{equation}
satisfies the momentum conservation
\begin{equation}
\begin{split}
q_{\rho}{\cal J}^{\rho}_{ss^{\prime}} &= (k_{\rho}+k_{\rho}^{\prime})\left( g^{\mu\sigma}\Bar{\psi}_{\mu}^{(+)s}(k)\gamma^{\rho}\psi_{\sigma}^{(-)s^{\prime}}(k^{\prime}) \right) \\
&= g^{\mu\sigma} \left( \Bar{\psi}_{\mu}^{(+)s}(k) \left( (\Slash{k}-m)+(\Slash{k}^{\prime}+m) \right) \psi_{\sigma}^{(-)s^{\prime}}(k^{\prime}) \right) \\
&= 0,
\end{split}
\label{normal1}
\end{equation}
where we used ${q}\equiv k+k^{\prime}$. $k$ and $k^{\prime}$ are the momentum of radiated particle and anti-particle, respectively. For convenience, we introduce the polarization tensors $P_{\mu\nu}^{(\pm)}$,
\begin{align}
P_{\mu \nu}^{(\pm)}(p)\equiv \sum_{s}\psi_{\mu}^{(\pm)s}(p)\Bar{\psi}_{\nu}^{(\pm)s}(p)=-(\Slash{p} \pm m)\left( g_{\mu\nu} -\frac{1}{3}\Pi_{\mu\nu}^{(\pm)}(p) \right) ,
\end{align} 
where
\begin{equation}
\Pi_{\mu\nu}^{(\pm)}(p) \equiv \gamma_{\mu}\gamma_{\nu} \pm \gamma_{\mu}\frac{p_{\nu}}{m} \mp \frac{p_{\mu}}{m}\gamma_{\nu} +2\frac{p_{\mu}p_{\nu}}{m^2}.
\end{equation}

Now we are ready to calculate the tree level amplitude of the Aharonov-Bohm radiation. Exploiting the gauge field around the AB string \eqref{ABsolution}, one obtain the amplitude 
\begin{equation}
i{\cal M}=i\frac{\epsilon}{4}\epsilon_{\sigma\rho\alpha\beta}{q^{\sigma}\over q_{\lambda}q^{\lambda}}{\cal J}_{ss^{\prime}}^{\rho}{J}^{\alpha\beta},
\end{equation}
where $\epsilon \equiv e\phi /\pi$ is the Aharonov-Bohm phase and $q$ is the momentum of the gauge field. In the momentum space, the string current ${J}^{\alpha \beta}$ is written as follows:
\begin{equation}
J^{\alpha \beta}= \int d\tau d \sigma \left( \dot{X}^{\alpha}X^{\prime \beta}-\dot{X}^{\beta}X^{\prime \alpha}  \right)
e^{iq\cdot X}.\label{newJ}
\end{equation}
A string configuration $X^{\mu}(\tau,\sigma)$ is a solution of wave equation, being composed of right movers and left movers,
\begin{equation}
X^{\mu}(\tau,\sigma)=\frac{1}{2}\left( a^{\mu}(\sigma - \tau )+b^{\mu}(\sigma +\tau )  \right),
\label{mover}
\end{equation}
where $a^0 = -\sigma_{-}=-(\sigma-\tau)$ and $b^0=\sigma_{+}=\sigma+\tau$. Plugging into \eqref{newJ}, one can rewrite ${J}^{\alpha\beta}$ in terms of $a,b, \sigma_{\pm}$ and the amplitude $i{\cal M}$ becomes 
\begin{equation}
i{\cal M} = i\frac{\epsilon}{4}\epsilon_{\sigma\rho\alpha\beta}{q^{\sigma}\over q_{\lambda}q^{\lambda}}{\cal J}^{\rho}_{ss^{\prime}} \left( I_{+}^{\alpha}I_{-}^{\beta}-I_{-}^{\alpha}I_{+}^{\beta} \right),
\end{equation}
where we defined the integrals,
\begin{equation}
\label{Ipm}
I_{+}^{\alpha}=\frac{1}{2}\int d\sigma_{+} \partial_{+}b^{\alpha}e^{iq\cdot b/2}, \qquad I_{-}^{\alpha}=\frac{1}{2}\int d\sigma_{-} \partial_{-}a^{\alpha}e^{iq\cdot a/2} ,
\end{equation}
where $\partial_{\pm}=\frac{\partial}{\partial \sigma_{\pm}}$.
These integrals satisfy the useful relations, $q\cdot I_{\pm}=0$. With these relations and \eqref{normal1}, the amplitude of the radiation can be further simplified as 
\begin{equation}
i{\cal M}=i\frac{\epsilon}{4q^0}\bm{ J}_{ss^{\prime}}\cdot (\bm{I}_{+}\times\bm{I}_{-}).
\end{equation}
Hence, the number of radiated spin 3/2 particles, we denote $dN$, with the momentum $k$ and $k^{\prime}$ is given by
\begin{equation}
dN=\frac{d^3k}{(2\pi)^3}\frac{1}{2k^0}\frac{d^3 k^{\prime}}{(2\pi)^3}\frac{1}{2k^{\prime 0}}\left( \frac{\epsilon}{4q^0} \right)^2\sum_{s,s^{\prime}} (\bm{I}_{+}\times \bm{I}_{-})_{i} (\bm{I}_{+}\times \bm{I}_{-})_{j}^{*} {\cal J}_{ss^{\prime}}^i ({\cal J}_{ss^{\prime}}^j)^*,
\label{dN}
\end{equation} 
where $i,j=1,2,3$. The sum of the spin polarization is done as follows:
\begin{equation}
\begin{split}
\sum_{s,s^{\prime}}{\cal J}^i_{ss^{\prime}} ({\cal J}^j_{ss^{\prime}})^* =& \sum_{s,s^{\prime}}g^{\alpha\beta}\left( \Bar{\psi}_{\alpha}^{(+)s}(k)\gamma^{i}\psi_{\beta}^{(-)s^{\prime}}(k^{\prime}) \right)
g^{\gamma\delta}\left( \Bar{\psi}_{\gamma}^{(+)s}(k)\gamma^{j}\psi_{\delta}^{(-)s^{\prime}}(k^{\prime}) \right)^{*} \\
=& g^{\alpha\beta}g^{\gamma\delta}\tr(P_{\gamma\alpha}^{(+)}(k)\gamma^{i}P_{\beta\delta}^{(-)}(k^{\prime})\gamma^{j})\\
= &
g^{\alpha\beta}g^{\gamma\delta}\tr[
(\Slash{k}+m)( g_{\gamma\alpha} -\frac{1}{3}\Pi_{\gamma\alpha}^{(+)}(k) ) \gamma^{i} (\Slash{k}^{\prime}-m)( g_{\beta\delta} -\frac{1}{3}\Pi_{\beta\delta}^{(-)}(k^{\prime}) ) \gamma^{j} ] \\
= & \frac{40}{9}(k^{i}k^{\prime j}+k^{\prime i}k^{j}-(m^2+k\cdot k^{\prime})g^{ij}) \\
&-\frac{16}{9}\left( k^{i}k^{j}+k^{\prime i}k^{\prime j} -(m^2 + k\cdot k^{\prime})g^{ij} \right) \\
&+\frac{16}{9}\frac{(k\cdot k^{\prime})^2}{m^4} \left( k^{i}k^{\prime j} +k^{\prime i}k^{j} -(m^2 + k\cdot k^{\prime})g^{ij} \right) ,
\end{split}\label{massiveformula}
\end{equation}
where we used $g^{\mu\nu}\Pi_{\mu\nu}^{(\pm)} = 6$ and $g^{\mu\lambda}g_{\lambda\nu}={\delta^{\mu}}_{\nu}$. This is the general formula for the AB radiation of spin $3/2$ fermions. The first term on the right hand side in \eqref{massiveformula} is the same contribution as the radiation of spin $1/2$ fermions \cite{ABrad2} up to overall numerical factor.

To go a step further, let us assume the string configuration. Along the lines of \cite{ABrad1,ABrad2}, we consider the cusp configuration, 
\begin{equation}
\begin{split}
b(\sigma_{+}) &= \frac{L}{2\pi}\left( \sin\left(\frac{2\pi \sigma_{+}}{L}\right),0,-\cos\left(\frac{2\pi \sigma_{+}}{L}\right) \right), \\
a(\sigma_{-}) &= \frac{L}{2\pi}\left( \sin\left(\frac{2\pi \sigma_{-}}{L}\right),-\cos\left(\frac{2\pi \sigma_{-}}{L}\right),0 \right) ,
\label{cusptra}
\end{split}
\end{equation}
and discuss the AB radiation of massive particle from the cusp. As is mentioned in the main text, the cusp (or kink) on a loop is the dominant radiation of a massive particle. Let us begin with calculations of $I^{\alpha}_{\pm}$. Using the periodic property of loops with size $L$, we can express $I^{\alpha}_{\pm}$ as discrete Fourier series expansions,
\begin{equation}
\begin{split}
I^{\alpha}_{+} &= 2\pi \sum_{l=-\infty}^{\infty} \delta(q^0+\frac{4\pi l}{L})\int_{0}^{L}\frac{d\sigma_{+}}{L}\partial_{+}b^{\alpha}e^{-il2\pi\sigma_{+}/L}e^{-i\bm{q}\cdot\bm{b}/2} ,\\
I^{\alpha}_{-} &= \frac{2\pi}{\delta_{\mathbb{Z}}(0)} \sum_{l=-\infty}^{\infty} \delta(q^0-\frac{4\pi l}{L})\int_{0}^{L}\frac{d\sigma_{-}}{L}\partial_{-}a^{\alpha}e^{-il2\pi\sigma_{-}/L}e^{-i\bm{q}\cdot\bm{a}/2},
\label{Ipm2} 
\end{split}
\end{equation}
where
\begin{equation}
\delta_{\mathbb{Z}}(0)\equiv \frac{\int_{-\infty}^{\infty}d\sigma e^{i2\pi l\sigma/L}}{\int_{0}^{L}d\sigma e^{i2\pi l \sigma/L}},\qquad  l\in \mathbb{Z}.
\end{equation}
Substituting \eqref{cusptra} in \eqref{Ipm2}, $\bm{I}_{+}\times \bm{I}_{-}$ becomes
\begin{equation}
\bm{I}_{+}\times \bm{I}_{-} = -\frac{16\pi^3}{L}\sum_{l=-\infty}^{\infty}\delta(q^0-\frac{4\pi l}{L})\bm{n_l},
\label{gaiseki}
\end{equation}
where we defined
\begin{equation}
\begin{split}
\bm{n_l} &\equiv (\bm{\hat{y}}\times \vec{\partial_{q}}\varphi_{(+|-l)})\times (\bm{\hat{z}}\times \vec{\partial_{q}}\varphi_{(-|l)}), \\
\varphi_{(+|l)}(q_x,q_z) &= i^l B_{l}\left( \frac{L}{4\pi}\sqrt{q_x^2+q_z^2} \right) \exp \left( il \arctan(q_x/q_z) \right), \\
\varphi_{(-|l)}(q_x,q_y) &= -i^l B_{l}\left( \frac{L}{4\pi}\sqrt{q_x^2+q_y^2} \right) \exp \left( il \arctan(q_x/q_y) \right) ,\\
\vec{\partial_{q}} &\equiv (\partial /\partial {q_x},\partial /\partial {q_y},\partial /\partial {q_z}).
\end{split}
\end{equation}
$B_{l}$ is the Bessel function. Applying the same strategy used in \cite{ABrad2}, we rewrite the squared $\delta$ function as follow:
\begin{equation}
\left( \delta(q^0-l\Omega) \right)^2 \rightarrow \delta(q^0-l\Omega)\frac{T}{2\pi}.
\label{delta1} 
\end{equation}
$l$ is integer, $\Omega$ is the characteristic frequency, and $T$ is the period. Integrating (\ref{dN}) with the relations (\ref{gaiseki}) and (\ref{delta1}), the rate of pair production $\dot{N}=N/T$ becomes
\begin{equation}
\begin{split}
\dot{N} &= \sum_{l}\int \frac{d^3 k}{k^0} \frac{d^3 k^{\prime}}{k^{\prime 0}} \frac{\epsilon^2}{36\pi L^2 (q^0)^2}\delta(q^0-\frac{4\pi l}{L}) \left[ \left( 5+2\frac{(k\cdot k^{\prime})^2}{m^4} \right) A_{1} -2A_{2} \right] \\
&\equiv \sum_{l}\dot{N}_{(l)},
\end{split}
\end{equation} 
where we defined
\begin{equation}
\begin{split}
A_{1} &= (m^2+k\cdot k^{\prime})|\bm{n_{l}}|^2+(\bm{n_l}\cdot \bm{k})(\bm{n_l}^{*}\cdot \bm{k^{\prime}}) + (\bm{n_l}\cdot \bm{k^{\prime}})(\bm{n_l}^{*}\cdot \bm{k})\\
A_{2} &= (m^2+k\cdot k^{\prime})|\bm{n_{l}}|^2+(\bm{n_l}\cdot \bm{k})(\bm{n_l}^{*}\cdot \bm{k}) + (\bm{n_l}\cdot \bm{k^{\prime}})(\bm{n_l}^{*}\cdot \bm{k^{\prime}}).
\end{split}
\end{equation}
Thus, the rate of radiated energy for the $l$-th mode, we denote $P_{(l)}$, is given by
\begin{equation}
\begin{split}
P_{(l)} &= q^{0}_{(l)}\dot{N}_{(l)} \\ 
&= \frac{\epsilon^2}{(12\pi)^2 lL}\int \frac{d^3 k}{k^0} \frac{d^3 k^{\prime}}{k^{\prime 0}} \delta(q^0-\frac{4\pi l}{L}) \left[ \left( 5+2\frac{(k\cdot k^{\prime})^2}{m^4} \right) A_{1} -2A_{2} \right],
\end{split}
\end{equation}
where $q^0_{(l)}=4\pi l/L$. This is the power of massive radiation of spin $3/2$ particles.
 
Finally, it is worthy emphasizing that in the massless limit, $m\to 0$, na\"ively, the terms in the last line of \eqref{massiveformula} diverge. However, in the limit, supersymmetry is restored, and the supercurrent, to which goldstino component of massive gravitino couples, becomes conserved. In this case, the terms in the last line of \eqref{massiveformula} do not contribute. Also, as is claimed in various papers (for example, see \cite{ZoharSeiberg, RychkovStrumia} for recent works), there is a coupling of the goldstino to the supercurrent ${\cal L}\sim  {1\over F} \bar{\psi}_{1/2}\partial_{\mu}S_{super}^{\mu}$. This interaction diverges in the limit $F\to 0$. In fact, the divergent terms in \eqref{massiveformula} come from the radiation of the goldstino originating from the coupling to the supercurrent. However as mentioned above, for the conserved supercurrent, this interaction vanishes and does not contribute. Remarkably, even if we ignore the divergent terms and taking the limit $m\to 0$, the result differs from the one for the exactly massless case. That is one of realizations of the known phenomena, van Dam-Veltman-Zakharov discontinuity \cite{Disc}. The goldstino coupling, $\gamma_{\mu} S^{\mu}_{super}$, which does not vanish even for the conserved supercurrent,  adds an extra contribution. Although the numerical order one factor is slightly different due to the extra contribution, as far as order estimation is concerned, applying the exactly massless case for evaluation is enough. Hence, in the main text, we study the exactly massless case for the sake of simplicity.

\setcounter{equation}{0}
\renewcommand{\theequation}{B.\arabic{equation}}
\section*{Appendix B\, Loop number density}

In this appendix, we briefly review the loop number density used in the main text. The loop number density of cosmic strings that loose the energy through two types of radiation was discussed in the literatures \cite{NewVacha2,VS,Vach,Dufaux,dilaton}. Among them, the discussion shown in \cite{NewVacha2} offers a transparent treatment of two kinds of emissions. So, we would like to review it with slight modifications of notation by focusing on AB and gravitational radiation. In general, the loop number density is written as
\begin{equation}
n(t,L)=\int_0^t dt_i  f(t_i, L_i) {\partial L_i \over \partial L}{a\over a_i} ,
\end{equation}
where $a(t)$ is the scale factor in the standard cosmology. The factor $f$ is the distribution function determined by dynamics of cosmic string network. From computer simulations \cite{BOS}, the function is expected as follows:
\begin{equation}
 f(t_i, L_i )\simeq  \left \{ 
 \begin{array}{l} 
 {\zeta_r t^{3/2}_i \sqrt{\alpha_r} \over L_i^{3/2}}\delta (\alpha_r t_i -L_i) \qquad \ \   ({\rm radiation \ dominant}) \\ 
 {\zeta_m t^{1.69}_i\sqrt{\alpha_m} \over L_i^{1.69}}\Theta (\alpha_m t_i -L_i) \qquad  ({\rm matter \ dominant})
 \end{array} 
 \right.  ,
\end{equation} 
where $\alpha_r \simeq \alpha_m  \simeq 0.1 \equiv \alpha$. The energy loss of cosmic strings is encoded in the factor ${\partial L_i \over \partial L}{a\over a_i} $. Hence, we need to know the behavior of the loop size $L$ under the evolution of the universe. The time dependence of the loop length is governed by the equation
\begin{equation}
\mu {d L\over d t}=- P_{AB} -P_{GW},
\end{equation}
where $P_{AB}$ and $P_{GW}$ are shown in \eqref{powerABrad} and \eqref{GWpower}. As is emphasized in \cite{NewVacha2}, it is quite involved to find a solution for this equation. However, at various stage of the universe, either gravitational or Aharonov-Bohm radiation is dominant. So, at the leading order, we can simply ignore the subdominant radiation at each era, and solve the equation approximately. In fact, the loop number density of cosmic strings with single radiation was already studied, 
\begin{equation}
 \left \{ 
 \begin{array}{l} 
 n_{GW}(t<t_{eq},L) = \frac{\zeta \sqrt{\alpha}  t^{-3/2}}{\left[ L+L_{GW} \right]^{5/2}} \Theta \left( 1-\frac{L}{\alpha t}\right) \\ 
 n_{GW}(t_{eq}<t,L) = \frac{\zeta \sqrt{\alpha}  t_{eq}^{1/2}t^{-2}}{\left[ L+L_{GW }\right]^{5/2}} \Theta(1-\frac{ L+L_{GW}-L_{GW}(t_{eq}) }{\alpha t_{eq}})  \\
 \qquad \qquad \qquad \qquad \qquad \qquad  \qquad \qquad  + \left[1-\left( {L \over 3 \beta_m t} \right)^{0.31} \right] {1 \over t^2} {\zeta_m \sqrt{\alpha} \over [L+L_{GW}]^2}
 \end{array} 
 \right. \label{nGW},
\end{equation} 
where $t_{ eq}\simeq 2.4\times 10^{12}{\rm [s]}$. $\beta_m\sim 0.06$ and the power $0.31$ in the experssion are estimated by computer simulations. The parameters $\zeta $ and $\zeta_m$, assumed to be ${\cal O}(10)$, are determined by dynamics of cosmic string network. $L_{GW}$ is defined by \eqref{minisize}. As for single AB radiation, 
\begin{equation}
 \left \{ 
 \begin{array}{l} 
 n_{AB}(t<t_{eq},L) = \frac{\zeta \sqrt{\alpha} L^{1/2} t^{-3/2}}{\left[ L^{3/2}+L_{AB}^{3/2} \right]^{2}} \Theta \left(1-\frac{L}{\alpha t }\right)  \\ 
 n_{AB}(t_{eq}<t,L) = \frac{\zeta \sqrt{\alpha} L^{1/2} t_{eq}^{1/2} t^{-2}}{\left[ L^{3/2}+L_{AB}^{3/2}\right]^{2}} \Theta \left(1-\frac{  L^{3/2}+L_{AB}^{3/2} -L_{AB}^{3/2} (t_{eq})}{(\alpha t_{eq})^{3/2} }\right) \\
 \qquad \qquad \qquad \qquad \qquad \qquad  \qquad \qquad  + \left[1-\left( {L \over 3\beta_m t} \right)^{0.31} \right] {L^{1/2} \over t^2} {\zeta_m \sqrt{\alpha}\over [L^{3/2}+L_{AB}^{3/2}]^{5/3}}
 \end{array} 
 \right.  \label{nAB}.
\end{equation} 
As discussed in \cite{NewVacha2}, by na\"ively adding two era, the total number density is given 
\begin{equation}
 n_{L} =  \left \{ 
 \begin{array}{l} 
 n_{GW}(t, L) \qquad  ({\rm GW\, rad \, dominant}) \\ 
  n_{AB}(t, L)\, \qquad  ({\rm AB\, rad \, dominant})
 \end{array} .
 \right.  \label{LoopAppendix}
\end{equation} 
Again, $L_{AB}$ is defined by \eqref{minisize}.

%
%

\end{document}